\documentclass[copyright,creativecommons]{eptcs}
\providecommand{\event}{Linearity 2016}
\usepackage{underscore}           
\usepackage{amsmath}
\usepackage{bm}
\usepackage{proof}
\usepackage{xifthen}
\usepackage{xspace}
\usepackage{wrapfig}
\usepackage{floatflt}

\newcommand{\lltens}{\mathbin{\otimes}}
\newcommand{\llone}{\mathbf{1}}
\newcommand{\llplus}{\mathbin{\oplus}}

\newcommand{\llpar}{\mathbin{\wp}}
\newcommand{\llwith}{\mathbin{\&}}

\newcommand{\false}{f}
\newcommand{\true}{t}

\newcommand{\pand}{\mathbin{\wedge\kern-1.5pt^+}}
\newcommand{\ptrue}{\true^+}
\newcommand{\por}{\mathbin{\vee}}

\newcommand{\nimp}{\mathbin{\supset}}
\newcommand{\nfalse}{\false^-}
\newcommand{\nand}{\mathbin{\wedge\kern-1.5pt^-}}
\newcommand{\ntrue}{\true^-}

\newtheorem{theorem}{Theorem}

\newtheorem{example}[theorem]{Example}

\newcommand{\Ascr}{{\cal A}}

\newcommand{\Xscr}{{\cal X}}

\newcommand{\mgu}[1]{\hbox{mgu}(#1)}
\newcommand{\seqx}[3][\Xscr]{#1\,;\,#2\vdash #3}
\newcommand{\seq}[2]{#1\vdash #2}
\newcommand{\bang}{\mathop{!}}
\newcommand{\quest}{\mathop{?}}

\newcommand{\valuation}[1]{v(#1)}

\newcommand{\malleq}{MALL$\relax^=$\xspace}
\newcommand{\mumalleq}{$\mu\hbox{MALL}\relax^=$\xspace}

\newcommand{\x}{\bar{x}}

\renewcommand{\t}{\bar{t}}

\newcommand{\ppathv}{\hbox{\sl path}}

\newcommand{\one}[3]{{#1}\buildrel{#2}\over\longrightarrow{#3}}
\newcommand{\tup}[1]{\langle #1 \rangle}

\newcommand{\nat }{\hbox{\sl nat}\xspace}
\newcommand{\plus}{\hbox{\sl plus}\xspace}
\newcommand{\lessthan}{\hbox{\sl lt}\xspace}

\newcommand{\limp}{\mathbin{-\hspace{-0.70mm}\circ}}
\newcommand{\Adj}{\hbox{\sl Adj}}

\title{A Proof Theory for Model Checking: An Extended Abstract}
\author{   Quentin Heath
\institute{LIX, \'Ecole Polytechnique}
\institute{Palaiseau, France}
\and       Dale Miller
\institute{Inria Saclay and LIX, \'Ecole Polytechnique}
\institute{Palaiseau, France}}
\def\titlerunning{A Proof Theory for Model Checking: An Extended Abstract}
\def\authorrunning{Q. Heath and D. Miller}
\begin{document}
\maketitle

\begin{abstract}
While model checking has often been considered as a practical
alternative to building formal proofs, we argue here that the theory
of sequent calculus proofs can be used to provide an appealing
foundation for model checking.  Since the emphasis of model checking
is on establishing the truth of a property in a model, we rely on the
proof theoretic notion of additive inference rules, since such rules
allow provability to directly describe truth
conditions. Unfortunately, the additive treatment of quantifiers
requires inference rules to have infinite sets of premises and the
additive treatment of model descriptions provides no natural notion of
state exploration.  By employing a focused proof system, it is
possible to construct large scale, synthetic rules that also qualify
as additive but contain elements of multiplicative inference.  These
additive synthetic rules---essentially rules built from the
description of a model---allow a direct treatment of state
exploration.  This proof theoretic framework provides a natural
treatment of reachability and non-reachability problems, as well as
tabled deduction, bisimulation, and winning strategies.  
\end{abstract}

\section{Introduction}
\label{sec:intro}

Model checking was introduced in the early 1980's as a way to
establish properties about (concurrent) computer programs that were
hard or impossible to establish using traditional, axiomatic proof
techniques such as those describe by Floyd and Hoare \cite{emerson08bmc}.
In this extended abstract we show that model checking can be given a proof
theoretic foundation using the sequent
calculus of Gentzen \cite{gentzen35}, the linear logic
of Girard \cite{girard87tcs}, and a treatment of fixed points
\cite{baelde12tocl,baelde07lpar,mcdowell00tcs,tiu12jal}.
The main purpose of this extended abstract
is foundational and conceptual.
Our presentation will not shed any new light on the algorithmic
aspects of model checking but it will show how model checkers can be
seen as having a ``proof search'' foundation shared with logic
programming and (inductive) theorem proving. 

Since the emphasis of model checking is on establishing the truth of a
property in a model, a natural connection with proof theory is via the
use of \emph{additive} connectives and their inference rules.
We illustrate in Section~\ref{sec:prop} how the proof theory of
additive connectives naturally leads to the usual notion of
truth-table evaluation for propositional connectives.
Relying only on additive connectives, however, fails to provide an
adequate inference-based approach to model checking since it only
rephrases truth-functional semantic conditions and requires rules with
potentially infinite sets of premises.

The proof theory of sequent calculus contains additional inference
rules, namely, the \emph{multiplicative} inference rules which can be
used to encode much of the algorithmic aspects of model checking such as, for
example, those related to determining reachability and simulation
(or winning strategies).
In order to maintain a close connection between model checking and
truth in model, we shall put additive inference rules back in the
center of our framework but this time these rules will be 
additive \emph{synthetic} inference rules.
The synthesizing process will allow multiplicative connectives and
inference rules to appear \emph{inside} the construction of synthetic
rules but they will not appear \emph{outside} such synthetic rules.
The construction of synthetic inference rules will be governed by the
well established proof theoretic notions of \emph{polarization} and
\emph{focused proof systems} \cite{andreoli92jlc,girard91mscs}.

The connection between the proof theory based on such synthetic
inference rules and model checking steps is close enough that
certificates for both reachability and non-reachability as well as
bisimulation and non-bisimulation are representable as sequent calculus
proofs.

\section{The basics of the sequent calculus}
\label{sec:basics}

Let $\Delta$ and $\Gamma$ range over \emph{multisets}
of formulas.  
A \emph{sequent} is either one-sided, written $\seq{}{\Delta}$, or
two-sided, written $\seq{\Gamma}{\Delta}$ (we first consider two-sided
sequents in Section~\ref{sec:hypotheticals}).
Inference rules have one sequent as their conclusion and zero or more
sequents as premises.
We divide inference rules into three groups: the \emph{identity}
rules, the \emph{structural} rules, and the \emph{introduction} rules.
The following are the two structural rules and two identity rules we consider.
\[
  \infer[\hbox{weaken}]{\seq{}{B,\Delta}}{\seq{}{\Delta}}
  \qquad
  \infer[\hbox{contraction}]{\seq{}{\Delta,B}}{\seq{}{\Delta,B,B}}
  \qquad
  \infer[\hbox{initial}]{\seq{}{B,\neg B}}{}
  \qquad
  \infer[\hbox{cut}]{\seq{}{\Delta,_1\Delta_2}}
                    {\seq{}{\Delta_1,B}\quad\seq{}{\Delta_2,\neg B}}
\]
The negation symbol $\neg(\cdot)$ is used here not as a logical
connective but as a function that computes the negation normal form of
a formula.
The remaining rules of the sequent calculus are introduction
rules: for these rules, a logical connective has an occurrence in the
conclusion and does not have an occurrence in the premises.
(We shall see several different sets of introduction inference rules
shortly.)

When a sequent calculus inference rule has two (or more) premises,
there are two natural schemes for managing the side formulas (i.e.,
the formulas not being introduced) in that rule.  The following rules
illustrate these two choices for conjunction.
\[
  \infer{\seq{}{B\wedge C,\Delta}}
        {\seq{}{B,\Delta}\quad \seq{}{C,\Delta}}
  \qquad
  \infer{\seq{}{B\wedge C,\Delta_1,\Delta_2}}
        {\seq{}{B,\Delta_1}\quad \seq{}{C,\Delta_2}}
\]
The choice on the left is the \emph{additive} version of the rule:
here, the side formulas in the conclusion are the same in all the
premises.
The choice on the right is the \emph{multiplicative} version of the
rule: here, the various side formulas of the premises are accumulated
to be the side formulas of the conclusion.
Note that the cut rule above is an example of a multiplicative
inference rule.
A logical connective with an additive right introduction rule is also
classified as additive.  In addition, the de Morgan dual and the unit
of an additive connective are also additive connectives.
Similarly, a logical connective with a multiplicative
right-introduction rule is called multiplicative; so are its de Morgan
dual and their units.

The multiplicative and additive versions of inference rules are, in
fact, inter-admissible if the proof system contains weakening and
contraction.
In linear logic, where these structural rules are not available, the
conjunction and disjunction have additive versions $\llwith$ and
$\llplus$ and multiplicative versions $\lltens$ and $\llpar$,
respectively, and these different versions of conjunction and
disjunction are not provably equivalent.
Linear logic provides two \emph{exponentials}, namely the $\bang$ and
$\quest$, that permit limited forms of the structural rules for suitable
formulas.
The familiar exponential law $x^{n+m}=x^n x^m$ extends to the logical
additive and multiplicative connectives since $\bang(B\llwith
C)\equiv\bang B\lltens \bang C$ and $\quest(B\llplus
C)\equiv\quest B\llpar \quest C$.

While we are interested in model checking as it is practiced, we shall
be interested in only performing inference in classical logic.
One of the surprising things to observe about our proof theoretical
treatment of model checking is that almost all of it can be seen as
taking place within the proof theory of linear logic, a logic that
sits behind classical (and intuitionistic) logic.
As a result, the distinction between additive and multiplicative
connectives remains an important distinction for our framework.
Also, weakening and contraction will not be eliminated completely but
will be available for only certain formulas and in certain inference
steps (echoing the fact that in linear logic, these structural rules
can be applied to formulas annotated with exponentials).

\section{Additive propositional connectives}
\label{sec:prop}

Let $\Ascr$ be the set of formulas built from the propositional
connectives $\{\wedge,\true,\vee,\false\}$ (no propositional
constants included).
Consider the following small proof system involving one-sided sequents.
\[
  \infer{\seq{}{B_1\wedge B_2,\Delta}}
        {\seq{}{B_1,\Delta}\quad \seq{}{B_2,\Delta}}
  \qquad
  \infer{\seq{}{\true,\Delta}}{}
  \qquad
  \infer{\seq{}{B_1\vee B_2,\Delta}}
        {\seq{}{B_1,\Delta}}
  \qquad
  \infer{\seq{}{B_1\vee B_2,\Delta}}
        {\seq{}{B_2,\Delta}}
\]
Here, $\true$ is the unit of $\wedge$, and $\false$ is the unit of $\vee$.
Notice that $\vee$ has two introduction rules while $\false$ has none.
Also, $\true$ and $\wedge$ are de Morgan duals of $\false$ and $\vee$,
respectively.  
We say that the
multiset $\Delta$ is provable if and only if there is a proof of
$\seq{}{\Delta}$ using these inference rules.
Also, we shall consider no additional inference rules (that is, no
contraction, weakening, initial, or cut rules): this
inference system is composed only of introduction rules and all of
these introduction rules are for \emph{additive} logical connectives.

The following theorem identifies an important property of this purely
additive setting.
This theorem is proved by a straightforward induction on the structure
of proofs.

\begin{theorem}[Strengthening]
  \label{thm:strength}
  If $\Delta$ is a multiset of $\Ascr$-formulas and $\seq{}{\Delta}$
  then $\exists\; B\in\Delta$ such that $\seq{}{B}$.
\end{theorem}

This theorem shows that provability of purely additive
formulas is independent of their context.
It also establishs that the proof system is consistent, since the
empty sequent $\seq{}{\cdot}$ is not provable.

The following three theorems state that the missing inference rules of
weakening, contraction, initial, and cut are all admissible in this
proof system.  The first theorem is an immediate consequence of
Theorem~\ref{thm:strength}.  The following two theorems are proved,
respectively, by induction on the structure of formulas and by
induction on the structure of proofs.

\begin{theorem}[Weakening \& contraction admissibility]
  \label{thm:wc}
  Let $\Delta_1$ and $\Delta_2$ be multisets of $\Ascr$-formulas
  such that $\Delta_1$ is a subset of $\Delta_2$ (when viewed as sets).
  If $\seq{}{\Delta_1}$ is provable then $\seq{}{\Delta_2}$  is
  provable.
\end{theorem}

\begin{theorem}[Initial admissibility]
  \label{thm:init}
  Let $B$ be a $\Ascr$-formula.  Then $\seq{}{B,\neg B}$ is provable.
\end{theorem}

\begin{theorem}[Cut admissibility]
  \label{thm:cut}
  Let $B$ be an $\Ascr$-formula and let $\Delta_1$ and $\Delta_2$ be
  multisets of $\Ascr$-formulas.
  If both $\seq{}{B,\Delta_1}$ and $\seq{}{\neg B,\Delta_2}$ are
  provable, then there is a proof of $\seq{}{\Delta_1,\Delta_2}$.
\end{theorem}

These theorems lead to the following truth-functional semantics for
$\Ascr$ formulas: 
define  $\valuation{\cdot}$ as a mapping from $\Ascr$
formulas to booleans such that $\valuation{B}$ is $\true$ if $\seq{}{B}$ is
provable and is $\false$ if $\seq{}{\neg B}$ is provable.
Theorem~\ref{thm:init} implies that
$\valuation{\cdot}$ is always defined and Theorem~\ref{thm:cut}
implies that $\valuation{\cdot}$ is functional 
(does not map a formula to two different booleans).
The introduction rules describe this function 
\emph{denotationally}: e.g., $\valuation{A\wedge B}$ is the
truth-functional conjunction of $\valuation{A}$ and $\valuation{B}$
(similarly for $\vee$).

While this logic of $\Ascr$-formulas is essentially trivial, we will
soon introduce much more powerful additive inference rules: their
connection to truth functional interpretations (a la model checking
principles) will arise from the fact that their provability is not
dependent on other formulas in a sequent.

\section{Additive first-order structures}
\label{sec:fos}

We move to first-order logic by adding terms, equality on terms, and
quantification.

We shall assume that some \emph{ranked signature} $\Sigma$ of term
constructors is given: such a signature associates to every
constructor a natural number indicating that constructor's arity.
Term constants are identified with signature items given rank 0.  A
$\Sigma$\emph{-term} is a (closed) term built from only constructors
in $\Sigma$ and obeying the rank restrictions.  For example, if
$\Sigma$ is $\{a/0, b/0, f/1, g/2\}$, then $a$, $(f~a)$, and
$(g~(f~a)~b)$ are all $\Sigma$-terms.%
We shall consider only signatures for which there exist
$\Sigma$-terms: for example, the set $\{f/1, g/2\}$ is not a valid
signature. 
The usual symbols $\forall$ and $\exists$ will be used for the
universal and existential quantification over terms.
We assume that these quantifiers range over $\Sigma$-terms for some
fixed signature.
The arities of ranked signatures will often not be listed explicitly.

The equality and inequality of terms will be treated as (de Morgan
dual) logical connectives in the sense that their meaning is given by
the following introduction rules.
\[
\infer{\seq{}{t=t,\Delta}}{}
\qquad\qquad
\infer[\hbox{$t$ and $s$ differ}]{\seq{}{t\not= s,\Delta}}{}
\]
Here, $t$ and $s$ are $\Sigma$-terms for some ranked
signature $\Sigma$.

Consider (only for the scope of this section) the following two 
inference rules for quantification.  In these introduction rules,
$[t/x]$ denotes the capture-avoiding substitution.%
\[
\infer[\exists]{\seq{}{\exists x.B,\Delta}}{\seq{}{B[t/x],\Delta}}
\qquad
\infer[\hbox{$\forall$-ext}]
    {\seq{}{\forall x.B,\Delta}}
    {  \{~\seq{}{B[t/x],\Delta}~~|~~\Sigma\hbox{-term}~t~\} }
\]
Although $\forall$ and $\exists$ form a de Morgan dual pair, the rule
for introducing the universal quantifier is not the standard one used
in the sequent calculus (we will introduce the standard one later).
This rule, which is similar to the $\omega$-rule \cite{schwichtenberg77hml},
is an extensional approach to modeling quantification: 
a universally quantified formula is true if all instances of it are true.

Consider now the logic built with the (additive) propositional
constants of the previous section and with equality, inequality, and
quantifiers.
The corresponding versions of all four theorems in
Section~\ref{sec:prop} holds for this logic.
Similarly, we can extend the evaluation function for
$\Ascr$-formulas to work for the quantifiers:  in particular, 
$\valuation{\forall x. B x}=\bigwedge_t \valuation{B t}$ and 
$\valuation{\exists x. B x}=\bigvee_t \valuation{B t}$.
Such a result is not surprising, of course, since we have repeated
within inference rules the usual semantic conditions.  
The fact that these theorems hold indicates that the proof theory we
have presented so far offers nothing new over truth functional
semantics.
Similarly, this bit of proof theory offers nothing appealing to model
checking, as illustrated by the following example.

\begin{example}
  \label{ex:subset}
  Let $\Sigma$ contain the ranked symbols $z/0$ and $s/1$ and let us
  abbreviate the terms $z$, $(s~z)$, $(s~(s~z))$, $(s~(s~(s~z)))$, etc
  by {\bf 0}, {\bf 1}, {\bf 2}, {\bf 3}, etc.
  Let $A$ and $B$ be the set of terms $\{{\bf 0}, {\bf 1}\}$ and
  $\{{\bf 0}, {\bf 1}, {\bf 2}\}$, respectively.
  These sets can be encoded as the predicate expressions
  $\lambda x. x={\bf 0}\vee x={\bf 1}$ and $\lambda x. x={\bf 0}\vee
  x={\bf 1}\vee x={\bf 2}$.
  The fact that $A$ is a subset of $B$ can be denoted by
  the formula $\forall x. \neg(A\,x)\vee B\,x$ or, equivalently, as
  \[
   \forall x. (x\not={\bf 0}\wedge x\not={\bf 1})\vee x={\bf 0}\vee x={\bf 1}\vee x={\bf 2}
  \]
  Proving this formula requires an infinite number of premises of the
  form $(t\not={\bf 0}\wedge t\not={\bf 1})\vee t={\bf 0}\vee t={\bf
    1}\vee t={\bf 2}$.  Since each of these premises can, of course, be
  proved, the original formula is provable, albeit with an ``infinite
  proof''. 
\end{example}
While determining the subset relation between two finite sets is a
typical example of a model checking problem, one would not use the
above-mentioned inference rule for $\forall$ except in the extreme
cases where there is a finite and small set of $\Sigma$-terms.
As we can see, the additive inference rule for
$\forall$-quantification generally leads to ``infinitary proofs''
(an oxymoron that we now avoid at all costs).

\section{Multiplicative connectives}
\label{sec:hypotheticals}

Our departure from purely additive inference rules now seems forced
and we continue by adding multiplicative inference rules.

Our first multiplicative connective is the intuitionistic implication:
since the most natural 
treatment of this connective uses two-sided sequents, we make the move
away from the one-sided sequents that we have presented so far (see
Figure~\ref{fig:new}). 
Note that taking the two multiplicative rules of implication right
introduction and initial yields a proof system that violates the
strengthening theorem (Section~\ref{sec:prop}):
the sequent $\seq{}{p\nimp q,p}$ is provable while neither 
$\seq{}{p\nimp q}$  nor $\seq{}{p}$ are provable.

A common observation in proof theory is that the curry/uncurry
equivalence between $A\nimp B\nimp C$ and $(A\wedge B)\nimp C$ can be
mimicked precisely by the proof system: in this case, such precision
does not occur with the additive rules for conjunction but 
rather with the multiplicative version of conjunction.
To this end, we add the multiplicative conjunction 
$\pand$ and its unit $\ptrue$ and, for the sake of symmetry, we rename
$\wedge$ as $\nand$ and $\true$ to $\ntrue$.
(The plus and minus symbols are related to the polarization of logical
connectives that is behind the construction of synthetic connectives.)
These two conjunctions and two truth symbols are logically equivalent
in classical and intuitionistic logic although they are different in
linear logic where it is more traditional to write $\llwith$, $\top$,
$\lltens$, $\llone$ for $\nand$, $\ntrue$, $\pand$, $\ptrue$,
respectively.
The ``multiplicative false'' $\nfalse$ (written as $\perp$ in linear
logic) can be defined as $t\not=t$ (assuming that there is a
first-order term $t$).

Eigenvariables are binders at the sequent level that
align with binders within formulas (i.e., quantifiers).
Binders are an intimate and low-level feature of logic: the addition
of eigenvariables requires redefining the notions of term and sequent.

\begin{figure}
\begin{footnotesize}
  \[
    \infer{\seqx{\Gamma}{A\nand B,\Delta}}
          {\seqx{\Gamma}{A,\Delta}\quad \seqx{\Gamma}{B,\Delta}}
    \qquad
    \infer{\seqx{\Gamma}{\ntrue,\Delta}}{}
    \qquad
    \infer{\seqx{\Gamma,A\nand B}{\Delta}}
          {\seqx{\Gamma,A}{\Delta}}
    \qquad
    \infer{\seqx{\Gamma,A\nand B}{\Delta}}
          {\seqx{\Gamma,B}{\Delta}}
  \]
  \[
    \infer{\seqx{\Gamma,A\vee B}{\Delta}}
          {\seqx{\Gamma,A}{\Delta}\quad \seqx{\Gamma,B}{\Delta}}
    \qquad
    \infer{\seqx{\Gamma,\false}{\Delta}}{}
    \qquad
    \infer{\seqx{\Gamma}{A\vee B,\Delta}}
          {\seqx{\Gamma}{A,\Delta}}
    \qquad
    \infer{\seqx{\Gamma}{A\vee B,\Delta}}
          {\seqx{\Gamma}{B,\Delta}}
  \]
  \[
    \infer{\seqx{\Gamma,\Gamma'}{A\pand B,\Delta,\Delta'}}
          {\seqx{\Gamma}{A,\Delta}\quad \seqx{\Gamma'}{B,\Delta'}}
    \quad
    \infer{\seqx{}{\ptrue,}}{}
    \qquad
    \infer{\seqx{\Gamma,A\pand B}{\Delta}}
          {\seqx{\Gamma,A,B}{\Delta}}
  \qquad
    \infer{\seqx{\Gamma,\ptrue}{\Delta}}
          {\seqx{\Gamma}{\Delta}}
  \]
  \[
    \infer{\seqx{\Gamma}{A\nimp B,\Delta}}{\seqx{\Gamma,A}{B,\Delta}}
    \qquad
    \infer{\seqx{\Gamma,\Gamma',A\nimp B}{\Delta,\Delta'}}
          {\seqx{\Gamma}{A,\Delta}\quad\seqx{\Gamma',B}{\Delta'}}
  \]
  \[
  \infer[\exists]{\seqx{\Gamma}{\exists x.B,\Delta}}{\seqx{\Gamma}{B[t/x],\Delta}}
  \qquad
  \infer[\forall]{\seqx{\Gamma}{\forall x.B,\Delta}}{\seqx[\Xscr,y]{\Gamma}{B[y/x],\Delta} }
  \qquad
  \infer{\seqx{}{t=t}}{\strut}
  \qquad
  \infer{\seqx{t\not=t}{}}{}
  \]
  \[
  \vbox{\hsize 5.5truecm \noindent When $t$ and $s$ are not unifiable,}
  \quad
  \infer{\seqx{\Gamma,t = s}{\Delta}}{}
  \qquad
  \infer{\seqx{\Gamma}{t\not= s,\Delta}}{}
  \]
  \[
  \vbox{\hsize 5.5truecm \noindent Otherwise, set $\theta=\mgu{t,s}$}
  \quad
  \infer{\seqx{\Gamma,t=s}{\Delta}}
        {\seqx[\theta\Xscr]{\theta\Gamma}{\theta\Delta}}
  \qquad
  \infer{\seqx{\Gamma}{t\not=s,\Delta}}
        {\seqx[\theta\Xscr]{\theta\Gamma}{\theta\Delta}}
  \]
\end{footnotesize}
\vskip -10pt
  \caption{Introduction rules for propositional constants,
    quantifiers, and equality.   The $\exists$ rule is
    restricted so that $t$ is a $\Sigma(\Xscr)$-term and the $\forall$
    rule is restricted so that $y\not\in\Xscr$.
}
  \label{fig:new}
\end{figure}

Let the set $\Xscr$ denote \emph{first-order variables} and let
$\Sigma(\Xscr)$ denote all terms built from constructors in $\Sigma$
and from the variables $\Xscr$: in the construction of
$\Sigma(\Xscr)$-terms, variables act as constructors of arity 0.  
(We assume that $\Sigma$ and $\Xscr$ are disjoint.)  
A $\Sigma(\Xscr)$\emph{-formula} is one where all term constructors
are taken from $\Sigma$ and all free variables are contained in
$\Xscr$.
Sequents are now written as $\seqx{\Gamma}{\Delta}$: the
intended meaning of such a sequent is that the variables in the set
$\Xscr$ are bound over the formulas in $\Gamma$ and $\Delta$.
We shall also assume that formulas in $\Gamma$ and $\Delta$ are all
$\Sigma(\Xscr)$-formulas.
All inference rules are modified to account for this additional
binding: see Figure~\ref{fig:new}.
The variable $y$ used in the $\forall$ introduction rule is called, of
course, an eigenvariable.

The left introduction rules for equality in Figure~\ref{fig:new}
significantly generalizes the version involving only closed terms by
making reference to unifiability and to most general unifiers.
In the latter case, the domain of the substitution $\theta$ is a
subset of $\Xscr$, and the set of variables $\theta\Xscr$ is the result of
removing from $\Xscr$ all the variables in the domain of $\theta$ and
then adding in all those variables free in the range of $\theta$.
This treatment of equality was developed independently by
Schroeder-Heister \cite{schroeder-heister93lics} and Girard
\cite{girard92mail} and has been extended to include simply typed
$\lambda$-terms \cite{mcdowell00tcs}. 

While the use of eigenvariables in proofs allows us to deal with
quantifiers with finite proofs, that treatment is not directly
related to model theoretic semantics.
In particular, the strengthening theorem does not hold for this proof
system. 
As a result, obtaining a soundness and completeness theorem for
this logic is no longer trivial.

The inference rules in Figure~\ref{fig:new} provide a
proper proof of the theorem considered in Example~\ref{ex:subset}.

\begin{example}
\label{ex:subset again}
Let $\Sigma$ and the sets $A$ and $B$ be as in
Example~\ref{ex:subset}.  Showing that $A$ is a subset of $B$ requires
showing that the formula $\forall x (A x\nimp B x)$ is provable.  That is,
we need to find a proof of the sequent
$
\seq{}{\forall x.(x={\bf 0}\vee x={\bf 1})
                  \nimp (x={\bf 0}\vee x={\bf 1}\vee x={\bf 2})}.
$
The following proof of this sequent uses the rules from
Figure~\ref{fig:new}: a double line means that two or more inference
rules might be chained together.
\begin{footnotesize}
\[
  \infer={\seqx[\cdot]{\cdot}{\forall x.(x={\bf 0}\vee x={\bf 1})
                  \nimp (x={\bf 0}\vee x={\bf 1}\vee x={\bf 2})}}{
    \infer{\seqx[x]{x={\bf 0}\vee x={\bf 1}}{x={\bf 0}\vee x={\bf 1}\vee x={\bf 2}}}{
      \infer{\seqx[x\kern-1pt]{x={\bf 0}}{x={\bf 0}\vee x={\bf 1}\vee x={\bf 2}}}{
        \infer={\seqx[\cdot]{\cdot}{{\bf 0}={\bf 0}\vee {\bf 0}={\bf 1}\vee {\bf 0}={\bf 2}}}{
          \infer{\seqx[\cdot]{\cdot}{{\bf 0}={\bf 0}}}{}}}
      &
      \infer{\seqx[x\kern-1pt]{x={\bf 1}}{x={\bf 0}\vee x={\bf 1}\vee x={\bf 2}}}{
        \infer={\seqx[\cdot]{\cdot}{{\bf 1}={\bf 0}\vee {\bf 1}={\bf 1}\vee {\bf 1}={\bf 2}}}{
          \infer{\seqx[\cdot]{\cdot}{{\bf 1}={\bf 1}}}{}}}}}
\]
\end{footnotesize}
The proof in this example is able to account for
a simple version of ``reachability'' in the sense that we only need to
consider checking membership in set $B$ for just those elements
``reached'' in $A$.
\end{example}

\section{Fixed points}
\label{sec:fixed}

A final step in building a logic that can start to provide a
foundation for model checking is the addition of least and greatest
fixed points and their associated rules for induction and coinduction.
Given that processes generally exhibit potentially infinite behaviors
and that term structures are not generally bounded in their size, it
is important for a logical foundation of model checking to allow for
some treatment of infinity.
The logic described by the proof system in
Figure~\ref{fig:new} is a two-sided version of \malleq (multiplicative
additive linear logic extended with first-order quantifiers and
equality) \cite{baelde07lpar}.
The decidability of this logic is easy to show: as one moves from
conclusion to premise in every inference rule, the number of
occurrences of logical connectives decrease.
As a result, it is a simple matter to write an exhaustive search
procedure that must necessarily terminate (such a search procedure can
also make use of the decidable procedure of first-order unification).

In order to extend the expressiveness of MALL, Girard added the
exponentials $\bang$, $\quest$ to MALL to get full linear logic
\cite{girard87tcs}.  The standard inference rules for exponentials
allows for some forms of the contraction rule
(Section~\ref{sec:basics}) to appear in proofs and, as a result,
provability is no longer decidable.  A different approach to extending
MALL with the possibility of having unbounded behavior was proposed in
\cite{baelde07lpar}: add to \malleq the least and greatest fixed point
operators, written as $\mu$ and $\nu$, respectively.  The proof
theory of the resulting logic, called \mumalleq, was been developed in
\cite{baelde12tocl} and exploited in a prototype model checker
\cite{baelde07cade}. 

Fixed point expressions are written as $\mu{}B\,\t$ or
$\nu{}B\,\t$, where $B$ is an expression representing a monotonic
higher-order abstraction, and $\t$ is a list of terms;
by monotonic, we mean that the higher-order argument of $B$ can only
occur in $B$ under even numbers of negations.
The unfolding of the fixed
point expressions $\mu B\,\t$ and $\nu B\,\t$ are $B(\mu{}B)\,\t$ and
$B(\nu{}B)\,\t$, respectively.

\begin{example}
  \label{ex:graph}
  Horn clauses (in the sense of Prolog) can be encoded as purely
  positive fixed point expressions.  For example, here is the Horn
  clause logic program (using the $\lambda$Prolog syntax, the
  \verb+sigma Y\+ construction encodes the quantifier $\exists{}Y$)
  for specifying a (tiny) graph and its transitive closure:
\begin{verbatim}
step a b.   step b c.    step c b.
path X Y :- step X Y.
path X Z :- sigma Y\ step X Y, path Y Z.
\end{verbatim}
  We can translate the \verb.step. relation into the binary predicate
  $\one{\cdot}{}{\cdot}$ defined by
  \begin{align*}
    \mu(\lambda{}A\lambda{}x\lambda{}y.\,
    (x=a\pand y=b)\por&(x=b\pand y=c)\por(x=c\pand y=b))
  \end{align*}
  which only uses positive connectives.  Likewise, \verb.path. can be
  encoded as the relation $\ppathv(\cdot,\cdot)$:
  \begin{align*}
    \mu(\lambda{}A\lambda{}x\lambda{}z.\,
    \one{x}{}{z}\por(\exists{}y.\,\one{x}{}{y}\pand{}A\,y\,z)).
  \end{align*}
To illustrate unfolding of the adjacency relation, note that unfolding
the expression $\one{a}{}{c}$ yields the formula
$    (a=a\pand c=b)\por(a=b\pand c=c)\por(a=c\pand c=b)$
which is not provable.  Unfolding the expression $\ppathv(a,c)$ and
performing $\beta$-reductions yields the expression
$    \one{a}{}{c}\por(\exists{}y.\,\one{a}{}{y}\pand{}\ppathv\,y\,c).$
\end{example}

\begin{figure}
\[
  \infer[\mu R]{\seqx{\Gamma}{\mu{}B\t,\Delta}}
                {\seqx{\Gamma}{B(\mu{}B)\t,\Delta}}
  \quad
  \infer[\mu L]{\seqx{\Gamma,\mu{}B\t}{\Delta}}
             {\seqx{\Gamma, S\t}{\Delta} &
              \seqx[\Xscr,\x]{BS\x}{S\x}}
\]
\[
  \infer[\nu L]{\seqx{\Gamma,\nu{}B\t}{\Delta}}
               {\seqx{\Gamma,B(\nu{}B)\t}{\Delta}}
  \quad
  \infer[\nu R]{\seqx{\Gamma}{\nu{}B\t,\Delta}}
               {\seqx{\Gamma}{S\t,\Delta} &
                \seqx[\x]{S\x}{BS\x}}
\]
\caption{Introduction rules for least ($\mu$) and greatest ($\nu$) fixed points}
\label{fig:fixed}
\end{figure}

In \mumalleq, both $\mu$ and $\nu$ are treated as logical connectives
in the sense that they will have introduction rules.  They are also de
Morgan duals of each other.  The inference rules for treating fixed
points are given in Figure~\ref{fig:fixed}.  The rules for induction
and coinduction ($\mu L$ and $\nu R$, respectively) use a higher-order
variable $S$ which represents the invariant and coinvariant in these
rules.  As a result, it will not be the case that cut-free proofs
will necessarily have the
sub-formula properties: the invariant and coinvariant are not
generally subformulas of the rule that they conclude.
The following unfolding rules are also admissible since they
can be derived using induction and coinduction.
\[
  \begin{array}{c}
  \infer{\seqx{\Gamma,\mu{}B\t}{\Delta}}
               {\seqx{\Gamma,B(\mu{}B)\t}{\Delta}}
  \qquad
  \infer{\seqx{\Gamma}{\nu{}B\t,\Delta}}
               {\seqx{\Gamma}{B(\nu{}B)\t,\Delta}}
  \end{array}
\]

The introduction rules in Figures~\ref{fig:new}
and~\ref{fig:fixed} are exactly the introduction rules of \mumalleq,
except for two shallow differences.  The first difference is that the
usual presentation of \mumalleq is via one-sided sequents (here, we
use two-sided sequents).  The second difference is that we have written
many of the connectives differently (hoping that our set of
connectives will feel more comfortable to those not familiar with
linear logic).  To be precise, to uncover the linear logic
presentation of formulas, one must translate
$\nand$, $\ntrue$, $\pand$,   $\ptrue$, $\vee$, and $\supset$ to
$\llwith$, $\top$, $\lltens$, $\llone$, $\oplus$, and $\limp$
\cite{girard87tcs}.  Note that the linear implication $B\limp C$ can
be taken as an abbreviation of $\neg B\llpar C$.

The following example shows that it is possible to prove some negations
using either unfolding (when there are no cycles in the resulting
state exploration) or induction.

\begin{example}
  \label{ex:negation}
Below is a proof that the node $a$ is not adjacent to $c$: the first
step of this proof involves unfolding the definition of the adjacency
predicate into its description.
\vskip-8pt
\begin{footnotesize}
\[
  \infer{\seq{\one{a}{}{c}}{\cdot}}{
  \infer{\seq{(a=a\pand c=b)\por(a=b\pand c=c)\por(a=c\pand c=b)}{\cdot}}{
     \infer{\seq{ a=a\pand c=b}{\cdot}}{\infer{\seq{a=a,c=b}{\cdot}}{}}
\quad
     \infer{\seq{ a=b\pand c=c}{\cdot}}{\infer{\seq{a=b,c=c}{\cdot}}{}}
\quad
     \infer{\seq{ a=c\pand c=b}{\cdot}}{\infer{\seq{a=c,c=b}{\cdot}}{}}}}
\]
\vskip -5pt
\end{footnotesize}
\noindent A simple proof exists for $\ppathv(a,c)$: one simply unfolds the fixed
point expression for $\ppathv(\cdot,\cdot)$ and chooses correctly when
presented with a disjunction and existential on the right of the
sequent arrow.
Given the definition of the path predicate, the following rules are
clearly admissible.  We write $\tup{t,s}\in\Adj$ whenever
$\seq{\cdot}{\one{t}{}{s}}$ is provable.
\[
\infer[\tup{t,s}\in\Adj]{\seqx{\Gamma,\ppathv(t,s)}{\Delta}}{\seqx{\Gamma}{\Delta}}
\qquad
\infer{\seqx{\Gamma,\ppathv(t,y)}{\Delta}}{\{\seqx{\Gamma,\ppathv(s,y)}{\Delta}\ |
\ \tup{t,s}\in\Adj\}}
\]
The second rule has a premise for every pair $\tup{t,s}$ of adjacent
nodes: if $t$ is adjacent to no nodes, then this rule has no premises
and the conclusion is immediately proved.  
A naive attempt to prove that there is no path from $c$ to $a$ gets
into a loop (using these admissible rules): attempt to prove 
$\seq{\ppathv(c,a)}{\cdot}$
leads to an attempt to prove
$\seq{\ppathv(b,a)}{\cdot}$
and again attempting to prove
$\seq{\ppathv(c,a)}{\cdot}$.
Such a cycle can be examined to yield an invariant that makes it
possible to prove the end-sequent.  In particular, the set of nodes
reachable from $c$ is $\{b,c\}$, subset of $N=\{a,b,c\}$.  The
invariant $S$ can be
described as the set which is the complement (with respect to $N\times
N$) of the set $\{b,c\}\times\{a\}$, or equivalently as the predicate
$\lambda x\lambda y. \bigvee_{\tup{u,v}\in S} (x=u\pand y=v)$.
With this invariant, the induction rule
($\mu L$) yields two premises.  The left premise simply needs to
confirm that the pair $\tup{c,a}$ is not a member of $S$.
The right premise sequent $\seqx[\x]{BS\x}{S\x}$ establishes that $S$
is an invariant for the $\mu{}B$ predicate.
In the present case, the argument list $\x$ is just a pair of
variables, say, $x,z$, and $B$ is the body of the $\ppathv$ predicate:
the right premise is the sequent
$
\ x,z\; ;\; \seq{\one{x}{}{z}\por(\exists{}y.\,\one{x}{}{y}\pand{}S\,y\,z)}{S\,x\,z}.
$
A formal proof of this follows easily by blindly applying applicable
inference rules.
\end{example}

While the rules for fixed points (via induction and coinduction) are
strong enough to transform cyclic behaviors into, for example,
non-reachabilty or (bi)simulation assertions,
these rules are not strong enough to prove
other simple facts about fixed points.  For
example, consider the following two named fixed point expressions used
for identifying natural numbers and the ternary relation of addition.
\begin{align*}
\nat =  &\mu\lambda N\lambda n(n=z\vee\exists n'(n=s n'\pand N~n'))\\
\plus = &\mu\lambda P\lambda n\lambda m\lambda p
       ((n=z\pand m=p)\vee
        \exists n'\exists p'(n=s n'\pand p=s p'\pand P~n'~m~p'))
\end{align*}
The following formula (stating that the addition of two numbers is commutative)
\[
\forall n\forall m\forall p.
  \nat~n \supset \nat~m \supset \plus~n~m~p\supset\plus~m~n~p
\]
is not provable using the inference rules we have described.  The
reason that this formula does not have a proof is not because the
induction rule ($\mu L$ in Figure~\ref{fig:fixed}) is not strong
enough or that we are actually sitting inside linear logic: it is
because an essential feature of inductive arguments is missing.
Consider attempting a proof by induction that the property $P$ holds
for all natural numbers.  Besides needing to prove that $P$ holds of
zero, we must also introduce an arbitrary integer $j$ (corresponding
to the eigenvariables of the right premise in $\mu L$) and show that
the statement $P(j+1)$ reduces to the statement $P(j)$.  That is,
after manipulating the formulas describing $P(j+1)$ we must be able to
find in the resulting argument, formulas describing $P(j)$.  Up until
now, we have only ``performed'' formulas (by applying introduction
rules) instead of checking them for equality.  More specifically, while
we do have a logical primitive for checking equality of terms, the
proof system described so far does not have an equality for comparing
formulas.  As a result, some of the most basic theorems are not
provable in this system.  For example, there is no proof of
$\forall n.(\nat~n \supset \nat~n)$.

Model checking is not the place where we should
be attempting proofs involving arbitrary infinite domains: inductive theorem
provers are used for that.  If we restrict to finite domains,
however, proofs appear.  For example, consider the less-than binary
relation defined as
\begin{align*}
\lessthan = \mu \lambda L\lambda x\lambda y
  ((x = z \pand \exists y'. y = s y') \vee
   (\exists x'\exists y'. x = s x' \pand  y = s y'\pand L~x'~y'))
\end{align*}
The formula $(\forall n. \lessthan~n~{\bf 10}\supset \lessthan~n~{\bf 10})$
has a proof that involves generating all numbers
less than 10 and then showing that they are, in fact, all less than 10.
Similarly, a proof of the formula
$
  \forall n\forall m\forall p(
     \lessthan~n~{\bf 10}\supset \lessthan~m~{\bf 10}\supset
    \plus~n~m~p\supset \plus~m~n~p)
$
exists and consists of enumerating 100 pairs of numbers $\tup{n,m}$
and checking that the result of adding $n+m$ yields the same value as
adding $m+n$.

The full proof system for \mumalleq contains the cut rule and the 
following two initial rules.
\[
  \infer[\mu\,init]{\seqx{\mu{}B\t}{\mu{}B\t}}{}
  \qquad
  \infer[\nu\,init]{\seqx{\nu{}B\t}{\nu{}B\t}}{}
\]
The more general instance of the initial rule can be eliminated in
favor of these two specific instances.

\section{Conclusions}

Linear logic is usually understood as being an intensional logic whose
semantic treatments are quite remote from the simple model theory
consideration of first-order logic and arithmetic.
Thus, we draw the possibly surprising conclusions that the proof
theory of linear logic provides a suitable framework for model checking.
Many of the salient features of linear logic---lack of structural
rules, two conjunctions and two disjunctions, polarization---play
important roles in this framework.
The role of linear logic here seems completely different and removed
from, say, the use of linear logic to model multiset rewriting and
Petri nets \cite{kanovich95apal}:
we use it instead as \emph{the logic behind logic}.
In order to capture model checking, we need to deal with possibly
unbounded behaviors in specifications.
Instead of using the rule of contraction (which states, for example,
that the hypothesis $B$ can be repeated as the two hypotheses $B, B$)
we have used the theory of fixed points: there, unfolding replaces
$\mu B \t$ with $(B (\mu B) \t)$, thus copying the definition of $B$.
The use of fixed points also allows for the direct and natural
applications of the induction and coinduction principles.
In the full version of this paper, we show how a focused proof system
for \mumalleq can be used to describe large scale (synthetic) additive 
inference rules that are built from smaller scale inference rules that
may be multiplicative.

There can be several benefits for establishing and developing model
checking within proof theory.
One way to integrate theorem provers and model checkers would be to
allow them to exchange proof certificates in a common language of
formulas and proofs.  The logic of \mumalleq is close to the logic and
proofs used in some inductive theorem provers.
Also, linear logic is rich in duality.  Certain techniques used in
  model checking topics should be expected to dualize well.  For
  example, what is the dual notion for least fixed points of the
  notion of bisimulation-up-to?  What does predicate abstraction look
  like when applied to greatest fixed points?
Proof theory is a framework that supports rich abstractions, including
term-level abstractions, such as bindings in terms.  Thus, moving from
model checking using first-order terms to using simply typed
$\lambda$-terms is natural in proof theory: such proof
theoretic investigations of model checking over linguistic structures
including binders have been studied in \cite{miller05tocl} and have
been implemented in the Bedwyr system \cite{baelde07cade} which has
been applied to various model checking problems related to the
$\pi$-calculus \cite{tiu05fguc,tiu10tocl}.

\medskip
\noindent{\bf Acknowledgments.} We thank the reviewers of an earlier
draft of this abstract for their comments.  This work has been funded
by the ERC Advanced Grant Proof\kern 0.8pt Cert.

\end{document}